# The Resurgence of Trumponomics: Implications for the Future of ESG Investments in a Changing Political Landscape

**Innocentus Alhamis**

**Southern New Hampshire University**

**i.alhamis@snhu.edu**

---

**Dr. Innocentus Alhamis is a professor at the School of Business at Southern New Hampshire University.**

---

**Abstract**

Public policy shapes the economic landscape, influencing everything from corporate behavior to individual investment decisions. For Environmental, Social, and Governance (ESG) investors, these policy shifts can create opportunities and challenges as they navigate an ever-changing regulatory environment. The contrast between the Trump and Biden administrations offers a striking example of how differing political agendas can affect ESG investments. Trump's first term was marked by deregulation and policies favoring fossil fuels, which created an uncertain environment for sustainable investments. When Biden assumed office, his focus on climate action and clean energy reinvigorated the ESG sector, offering a more stable and supportive landscape for green investments. However, with Trump's return to power in his second term, these policies are being reversed again, leading to further volatility. This paper explores how such dramatic shifts in public policy influence economic strategies and directly impact ESG investors' decisions, forcing them to constantly reassess their portfolios in response to changing political climates.

**Introduction**

This paper examines the differing economic approaches of the Trump and Biden administrations, focusing on their impact on Environmental, Social, and Governance (ESG) investments. It looks at how Trump's focus on deregulation and fossil fuel-driven policies affected the ESG landscape, highlighting how changes in political leadership can shape trends in ESG and influence the future of sustainable investments.

**Trump's First Era Economic Vision: Trickle-Down Economics and Deregulation**

During his first term, President Donald Trump's economic agenda, often called "Trumponomics," was primarily centered on deregulation, tax cuts for businesses and individuals, and increasing domestic energy production. This approach mirrored the principles of Reaganomics, a trickle-down economic model that formed the foundation of Trump's economic policies (Tebaldi, 2020).

Trump's administration believed reducing taxes for corporations and the wealthiest individuals would spur investment and job creation, benefiting the broader population. However, critics argue that this approach disproportionately favors the wealthiest Americans and large corporations, increasing income inequality (Tebaldi, 2020).

Deregulation was a key focus of Trump's economic agenda, especially in the energy sector. During his four years in office, President Trump's administration rolled back significant climate policies and environmental regulations, impacting areas such as clean air, water, wildlife, and toxic chemicals. According to a New York Times analysis, supported by research from Harvard and Columbia Law Schools, nearly 100 environmental rules were either reversed or revoked under Trump, with more than a dozen potential rollbacks still in progress by the end of his term (Popovich, et al., 2021, January 20). Trump's "drill, baby, drill" approach emphasized fossil fuels, prioritizing oil and natural gas extraction over the global shift toward climate action (Sullivan, 2017).

**Bidenomics Approach to Economic Recovery and Sustainable Growth**

When President Joe Biden assumed office in January 2021, the United States was grappling with the fallout of the COVID-19 pandemic. The economy was in turmoil, and recovery seemed a daunting task. In response, Biden introduced a comprehensive economic strategy known as "Bidenomics", which provided a marked departure from the policies of his predecessor, Donald Trump.

Bidenomics contrasted sharply with Trump's economic agenda. While Trump's economic policy assumed that tax cuts for the wealthy and corporations would eventually benefit all by spurring investment and job creation, Bidenomics prioritized equitable economic growth and investing in people, the middle class, and underserved communities to ensure that society broadly shares prosperity (Biden, 2021).

At the heart of Bidenomics are three key priorities: public investment, empowerment of the middle class, and promotion of business competition. These goals aimed to foster long-term economic growth while addressing systemic inequalities and ensuring recovery benefits all Americans, particularly those historically left behind by previous economic policies (The White House, 2021).

Public investment was the cornerstone of Bidenomics. The administration directed substantial resources into modernizing infrastructure, creating clean energy jobs, and expanding access to healthcare. Key pieces of legislation, such as the American Rescue Plan, the Infrastructure Investment and Jobs Act, and the Inflation Reduction Act, have facilitated these investments. By focusing on clean energy and job creation, Bidenomics sought to build a sustainable and resilient economy (Biden, 2021).

Empowering the middle class was another critical aspect. Bidenomics prioritized increasing access to education, expanding vocational training, and strengthening unions. The administration focused on ensuring that workers have opportunities to climb the economic ladder, reduce inequality, and ultimately drive prosperity (The White House, 2021).

Lastly, Bidenomics promoted competition to benefit consumers and businesses. This included efforts to break up monopolies, particularly in the healthcare and tech sectors, to foster lower consumer prices and improve worker wages (Biden, 2021).

**Trump's Second Term Economic Vision: Reasserting Trickle-Down Economics**

During Donald Trump's inauguration for his second term as the president of the United States of America on 20th January 2025, President Trump reinstated his former policies and returned to familiar ground, focusing on deregulation and fossil fuel production. On his first day in office, he signed a series of executive orders to reverse Biden's initiatives. These included measures that undermined federal workers' rights and signaled upcoming cuts to federal programs by establishing the Department of Government Efficiency (Chowdhury et al 2025, January 22).

One of the most significant moves was Trump's executive order to withdraw the U.S. from the Paris Climate Agreement, distancing the country from global efforts to reduce greenhouse gas emissions. He also reversed clean energy initiatives and environmental protections and prioritized fossil fuel interests.

Trump further sought to reduce regulatory burdens, particularly in electric vehicles and energy efficiency standards. His administration aimed to streamline energy project approvals and remove perceived red tape. Additionally, Trump lifted the moratorium on new liquefied natural gas (LNG) export licenses, directing the Department of Energy to resume processing applications for new permits, reversing Biden's pause in 2024.

**Bidenomics and ESG Investments**

The stark differences between Trump's and Biden's policies and their resulting outcomes highlight the critical role that public policies play in shaping the regulatory landscape for businesses and influencing investor decisions.

Bidenomics significantly shaped the ESG investment landscape by prioritizing sustainability, social equity, and strong governance.

A key aspect of Bidenomics is the government's emphasis on green energy. This is reflected in the administration's policies to reduce carbon emissions, support renewable energy, and re-engage with international climate agreements, such as the Paris Climate Accord.

The Biden administration invested heavily in renewable energy technologies, including wind and solar power, to transition the U.S. to a cleaner energy future. The Biden administration set the stage for a robust green energy economy through significant legislative actions such as the American Rescue Plan, the Infrastructure Investment and Jobs Act, and the Inflation Reduction Act. These investments were essential for ESG investors, who are increasingly seeking opportunities in sectors that align with environmental sustainability goals (Biden, 2021).

The Inflation Reduction Act included provisions for clean energy incentives, electric vehicle production, and developing a clean energy workforce (Biden, 2022). These policies signal to the market that the U.S. government was committed to supporting sustainable industries, positively impacting ESG investment trends. By creating a favorable regulatory environment, Bidenomics encouraged capital flow into green energy sectors, which investors viewed as both environmentally responsible and economically promising long-term (Clark, 2015).

Bidenomics also integrated social and governance factors by expanding access to education, improving healthcare, and supporting workers' rights. The administration pushed for greater union participation, fair wages, and economic opportunities for underserved communities. ESG investors, who prioritize the "S" (social) and "G" (governance) factors, found these policies attractive, as they align with the principles of inclusivity, fairness, and human rights. Biden's support for expanding healthcare access and his focus on reducing inequality enhanced the social dimension of ESG investing (The White House, 2021).

In addition, the Biden administration has taken steps to improve corporate governance by advocating for increased transparency and accountability. Policies that promote good governance, including anti-corruption measures and better labor rights protections, offer a strong framework for ESG investors seeking to support companies with robust governance practices (Biden, 2021).

Under the Biden administration, a clear shift toward policies that foster long-term sustainability had been made. By reversing the previous administration's rollback of environmental regulations, the Biden administration provided more certainty for investors in clean energy and other sustainable sectors (Friede et al., 2015). This stability was crucial for ESG investors who need to assess the risks and rewards of their investments based on long-term policy trends. The government's alignment with international climate goals, such as the Paris Agreement, further reassures investors that the U.S. is committed to comprehensively addressing climate change (Biden, 2021).

**The Impact of Policy Changes on ESG Investments**

Trump's policies, which focus on reducing dependence on foreign energy and promoting economic growth through deregulation, are designed to open opportunities for investors in the oil and gas sectors. These policies are framed as necessary to ease the burden on businesses and stimulate growth, particularly in fossil fuel industries. By prioritizing oil, gas, and coal production, Trump's agenda creates investment opportunities in these sectors (Tebaldi, 2020). However, in the long term, the rollback of environmental protections and clean energy policies may create volatility for ESG investors, as they encounter an unpredictable regulatory landscape that could undermine the stability of green investments.

In contrast, Biden's public policies provided investors with a solid legal, financial, and informational infrastructure to align their capital with sustainable and socially responsible practices. Through regulation, incentives, transparency, and global coordination, Biden's policies fostered a climate where ESG considerations could be integrated into investment strategies, promoting a more sustainable and equitable economic future (Clark, 2015). Overall, policies that support green energy and sustainable practices offer a clear framework for investors looking to align their capital with long-term environmental goals. However, when political leadership rolls back these policies, market uncertainty increases, making it harder for ESG investors to assess the risks and rewards of their investments (Clark, 2015).

The Trump administration's withdrawal from the Paris Agreement and its scaling back of climate initiatives likely heightened uncertainty and instability in the clean energy sector. This left investors navigating a more unpredictable regulatory landscape (Friede et al., 2015). For ESG investors, the tension between political leadership and sustainability goals remains a key challenge. When policies prioritize fossil fuel production and downplay climate change, investors must reassess their portfolios and adapt to shifting regulatory environments. This unpredictability, especially during Trump's tenure, complicates the ESG investment landscape, deterring long-term commitments to green investments.

Public policy shifts from supporting to undermining environmental, social, and governance (ESG) factors often result in sudden changes to the regulatory framework, financial incentives, and market dynamics, making ESG investments less attractive or riskier.

Under Trump, ESG investors who had placed their capital in renewable energy sectors based on the assumption of favorable regulatory frameworks now faced a less supportive policy environment. The shift in regulations made it harder for investors to forecast the trajectory of green energy investments as the previously favorable market conditions became uncertain.

Following Biden's push for climate action, the second Trump administration again discourages ESG investors, especially with its approach to renewable energy. Reducing or eliminating financial incentives for clean energy projects presents a tough challenge. For instance, the corporate tax cuts from Trump's 2017 Tax Cuts and Jobs Act lowered tax rates for businesses, making clean energy tax incentives far less attractive, like those for wind and solar projects. This shift in policy reduced the financial support available for renewable energy projects, making them less appealing to investors. In contrast, Biden's administration focused heavily on climate change, introducing policies like the Inflation Reduction Act that offered generous tax credits and incentives for clean energy investments, creating a more supportive financial environment for ESG investors. However, with the return of policies that favor fossil fuels and reduce clean energy incentives, ESG investors once again face a more uncertain and discouraging landscape.

The shifting market perception and investor confidence in ESG-related sectors also play a crucial role. Policy changes can alter corporate behavior, especially regarding sustainability commitments. During the first Trump administration, the U.S. withdrawal from the Paris Agreement signaled a reduced commitment to global climate goals, affecting market expectations. This created an environment where ESG investments are less attractive, particularly in sectors aligned with environmental sustainability.

When policies constantly shift back and forth, ESG investors feel uncertain about where to place their trust. For example, Biden's re-entry into the Paris Agreement and emphasis on climate action helped restore confidence in ESG sectors, driving investments in green technologies and renewable energy. However, when Trump pulled out of the Paris Agreement again, it threw investors off balance, making it harder for them to commit to their plans and questioning whether they should continue supporting ESG initiatives. The back-and-forth nature of these policy changes creates a sense of instability, making long-term commitment difficult for those looking to invest in sustainable solutions.

Under Biden's administration, companies were encouraged to invest in cleaner energy solutions and improve their ESG performance, making these investments more appealing again. However, when Trump returned to office and rolled back environmental regulations, the shift in policy created a stark contrast. With fewer restrictions on fossil fuel industries, companies are suddenly incentivized to prioritize short-term profits over sustainable practices. This undermines the efforts of ESG investors trying to support businesses that focus on environmental or social responsibility, leaving them feeling demoralized and questioning whether their investments align with their values in such an unpredictable landscape.

**Conclusion**

The return of Trumponomics, which emphasizes deregulation and fossil fuel production, starkly contrasts Bidenomics' focus on sustainability and green energy. This ongoing clash between short-term economic growth strategies and long-term sustainability goals presents significant challenges for ESG investors. As political leadership shifts, so does the landscape for sustainable investments. For those seeking to align their portfolios with environmental goals, navigating this evolving political and regulatory terrain will be crucial for ensuring the continued growth of sustainable finance.

**Bibliography**

Biden, J. (2021). Building a Better America: Bidenomics and Economic Renewal. The White House. Retrieved from https://www.whitehouse.gov

Biden, J. (2022). Inflation Reduction Act. The White House. Retrieved from https://www.whitehouse.gov

Chowdhury, M., Shelton, S., Shen, M., Hammond, E., & Sangal, A. (2025, January 22). Trump seeks to reshape the US government with sweeping executive actions. Retrieved from https://www.cnn.com/politics/live-news/trump-president-executive-actions-01-22-25/index.html

Clark, G. L. (2015). The Origins of Sustainable Finance. In Handbook of Research on Sustainable Finance. Edward Elgar Publishing.

Friede, G., Busch, T., & Bassen, A. (2015). ESG and financial performance: aggregated evidence from more than 2000 empirical studies. Journal of Sustainable Finance & Investment, 5(4), 210-233. https://doi.org/10.1080/20430795.2015.1118917